\documentclass[aps,prb,reprint]{revtex4-2}
\usepackage{amsmath}
\usepackage{graphicx}
\usepackage{amssymb}
\usepackage{dcolumn}
\usepackage{multirow}

\begin{document}
\title{Stacking dependent ferroelectricity and antiferroelectricity in quasi-one-dimensional oxyhalides NbO$X_3$}
\author{Wencong Sun$^1$}
\author{Ning Ding$^1$}
\author{Jun Chen$^1$}
\author{Hai-Peng You$^2$}
\author{Jin Peng$^1$}
\author{Shan-Shan Wang$^1$}
\email{Email: wangss@seu.edu.cn}
\author{Shuai Dong$^1$}
\email{Email: sdong@seu.edu.cn}

\affiliation{$^1$School of Physics, Southeast University, Nanjing 211189, China}
\affiliation{$^2$School of Science, Changzhou Institute of Technology, Changzhou 213032, China}

\date{\today}

\begin{abstract}
Low-dimensional ferroelectricity and polar materials have attracted considerable attentions for their fascinating physics and potential applications. Based on first-principles calculations, here we investigate the stacking modes and polar properties of a typical series of quasi-one-dimensional ferroelectrics: double-chain oxyhalides NbO$X_3$ ($X$=Cl, Br, and I). The geometry of their double-chains allows both the interchain/intrachain permutation. Thus, different stacking modes of double-chains lead to a variety of ferroelectric and antiferroelectric phases in both the tetragonal and the monoclinic crystals. The proximate energies of these phases may lead to multiphase coexistence in real materials, as well as the hydrostatic pressure driving structural phase transition. Their spontaneous polarizations and piezoelectricity of the ferroelectric phases are prominent, comparable to commercially used ferroelectric BaTiO$_3$ and piezoelectric ZnO, respectively. Our paper demonstrates that the van der Waals NbO$X_3$ are promising materials for exploring quasi-one-dimensional ferroelectricity and antiferroelectricity.
\end{abstract}
\maketitle

\section{Introduction}
Ferroelectrics with spontaneous electric polarizations have stimulated great interests in condensed matter and material communities, due to their significant scientific importance and wide applications in devices \cite{rabe2007modern,Dawber2005,xue2021emerging}. Especially, the piezoelectric properties of ferroelectrics enable the interconversion between force/pressure and electrical signals, leading to various electromechanical sensors and actuators \cite{zhou2021van,Wang2007,briscoe2014piezoelectricity}. Especially, the intrinsic piezoelectric coefficients of ferroelectric (FE) materials are significantly enhanced near the morphotropic phase boundary (MPB) between two proximate phases with similar crystallographic structures but different polarities \cite{Noheda2000}. Most studies on MPB's have been performed on ferroelectric perovskites, such as Pb(Zr$_x$Ti$_{1-x}$)O$_3$ and (PbMg$_{1/3}$Nb$_{2/3}$O$_3$)$_{1-x}$–(PbTiO$_3$)$_x$ \cite{ye2008handbook,Guo2000}. Recently, MPB's have also been identified or predicted in simpler compounds, such as BaTiO$_{3}$, PbTiO$_{3}$ \cite{Wu2005a,Fu2000}, as well as a GeS monolayer \cite{Song2021}.

When FE-related devices are reduced to nanoscale, traditional three-dimensional (3D) FE materials may become incompetent, since their ferroelectric properties are usually severely suppressed due to the depolarization field and surface reconstruction with dangling bonds \cite{Dawber2005}. The rise of low-dimensional materials opens up a new platform for searching ferroelectrics for next generation micro-electro-mechanical systems \cite{an2020ferroic,wu2021two,guan2020recent,You2020,Wang2021}. Compared with traditional 3D counterparts, low-dimensional materials are easier to be manipulated, naturally suitable for the miniaturized devices. The study of two-dimensional polar systems started from the graphene monolayer functionalized by hydroxyl groups \cite{Wu2013a,Kan2013}. Later, intrinsic ferroelectricity has also been achieved or predicted in many monolayers or few-layers, most of which are exfoliated from van der Waals (vdW) materials \cite{Ding2017,Liu2018a,Bruyer2016,Fei2016a,Guan2018,Wu2016a,Xiao2018,Ding2021}.

The dimension of FE materials can be further reduced to one-dimensional (1D). For example, organic PVDF-TrFE nanowire \cite{Hu2009}, group-IV metal chalcogenides nanowires \cite{Zhang2019}, and BiN/SbN nanowires \cite{Yang2021} were confirmed/predicted to exhibit ferroelectricity. In addition, quasi-1D ferroelectricity was also predicted in vdW oxyhalides WO$X_4$ ($X$=Cl, Br, and I) \cite{Lin2019}. The 1D antiferroelectric (AFE) materials are also predicted, such as CsTaS$_{3}$ \cite{Bie2022}. Even though, the known 1D FE/AFE materials remain rare, and more seriously the experimental realization of these 1D polar materials remains challenging. Therefore, it is urgent to explore more suitable (quasi)-1D FE/AFE materials, especially those can be easily synthesized and manipulated. 

NbO$X_3$ ($X$=Cl, Br, and I) bulks with vdW interactions between 1D double-chains were experimentally synthesized many years ago \cite{Sands1959,Strobele2002,Hartwig2008a}, which were found to own polar structures. Recently, Zhang \textit{et al.} investigated the FE properties of the individual double-chain of NbO$X_3$ \cite{Zhang2021a}, confirming their ferroelectricity in the 1D limit. Thus, as an experimentally-existing series, this Nb-based oxyhalide family is highly important as a prototype model system to explore the 1D ferroelectricity. However, there are still many unresolved physical issues for NbO$X_3$, especially for their bulk structures. The geometry of vdW double-chains allows different configurations of FE dipoles, leading to different polar phases. In Ref.~\cite{Zhang2021a}, the intra-chain AFE and FE configurations were considered, whereas the inter-chain configurations (including the chain stacking and dipole alignments) were not touched.

In this paper, we thoroughly investigate the stability and polarity of NbO$X_3$ bulk structures using first-principles calculations. The different stacking patterns of these 1D double-chains lead to the presence of seven possible polar phases within the tetragonal and monoclinic frameworks. Our calculation confirms that the displacements of Nb ions break the local inversion symmetry, which lead to robust ferroelectricity or antiferroelectricity in bulks. The proximate energies of FE and AFE phases allow possible MPB's in these systems, whereas the original experimental x-ray diffraction (XRD) patterns might miss the FE phase. Furthermore, a moderate hydrostatic pressure can tune the phase transitions between the tetragonal and the monoclinic structures.

\section{Computational Methods}
Our first-principles density functional theory (DFT) calculations are carried out using the Vienna {\it ab initio} simulation package in the framework of density functional theory \cite{kresse1996}. The projector augmented wave method is adopted to deal with the interaction between electrons and ions \cite{Blochl1994}. The exchange-correlation functional is modeled by the generalized gradient approximation parameterized by Perdew, Burke, and Ernzerhof (PBE) \cite{PerdewPBE}. Although Nb is a transition metal ion, its $4d$ orbitals are empty in the Nb$^{5+}$ state. Thus, the Hubbard $U$ repulsion has little influence on the physical properties of NbO$X_3$. Therefore, our calculations are based on the pure PBE method, as performed for WO$X_4$ \cite{Lin2019}.

The energy cutoff for the plane-wave basis is set to be $520$ eV. The Monkhorst-Pack $k$-mesh of $3\times3\times9$ is adopted for the Brillouin zone (BZ) sampling for both the tetragonal and monoclinic lattices. The lattice constants and ionic positions are fully optimized until the residual force on each atom is less than $0.005$ eV/\AA. The energy convergence criterion is set to be $10^{-6}$ eV. Besides, to accurately describe the vdW interaction between double-chains, zero damping DFT-D3 of Grimme correction is applied \cite{Grimme2010}. 

Dynamical stability is demonstrated by phonon spectral calculations based on finite differences and PHONOPY \cite{Togo2015}. The phonon modes are analyzed by the AMPLIMODES software \cite{2009AMPLIMODESSA}. The ferroelectric polarization is calculated using the Berry phase method \cite{Resta1994,Berryphase2}. The piezoelectric coefficient is calculated based on via finite differences \cite{LePage2002}.

In recent years, machine learning interatomic potentials becomes more and more popular to predict material structures and properties \cite{MLIP1,MLIP2,MLIP3}, which can be more efficient than conventional DFT, especially for those large systems. However, the systems involved here are unit cells (u.c.s) with only 20 ions. Then the DFT calculation is more suitable to deal with the subtle balance of candidate phases.

Besides the ground state, the thermal stability has also been checked using the \textit{ab-initio} molecular dynamics (AIMD) simulations.

\section{Result and Discussion}
\subsection{Structures of bulk NbO$X_3$}
\begin{figure}
\includegraphics[width=0.47\textwidth]{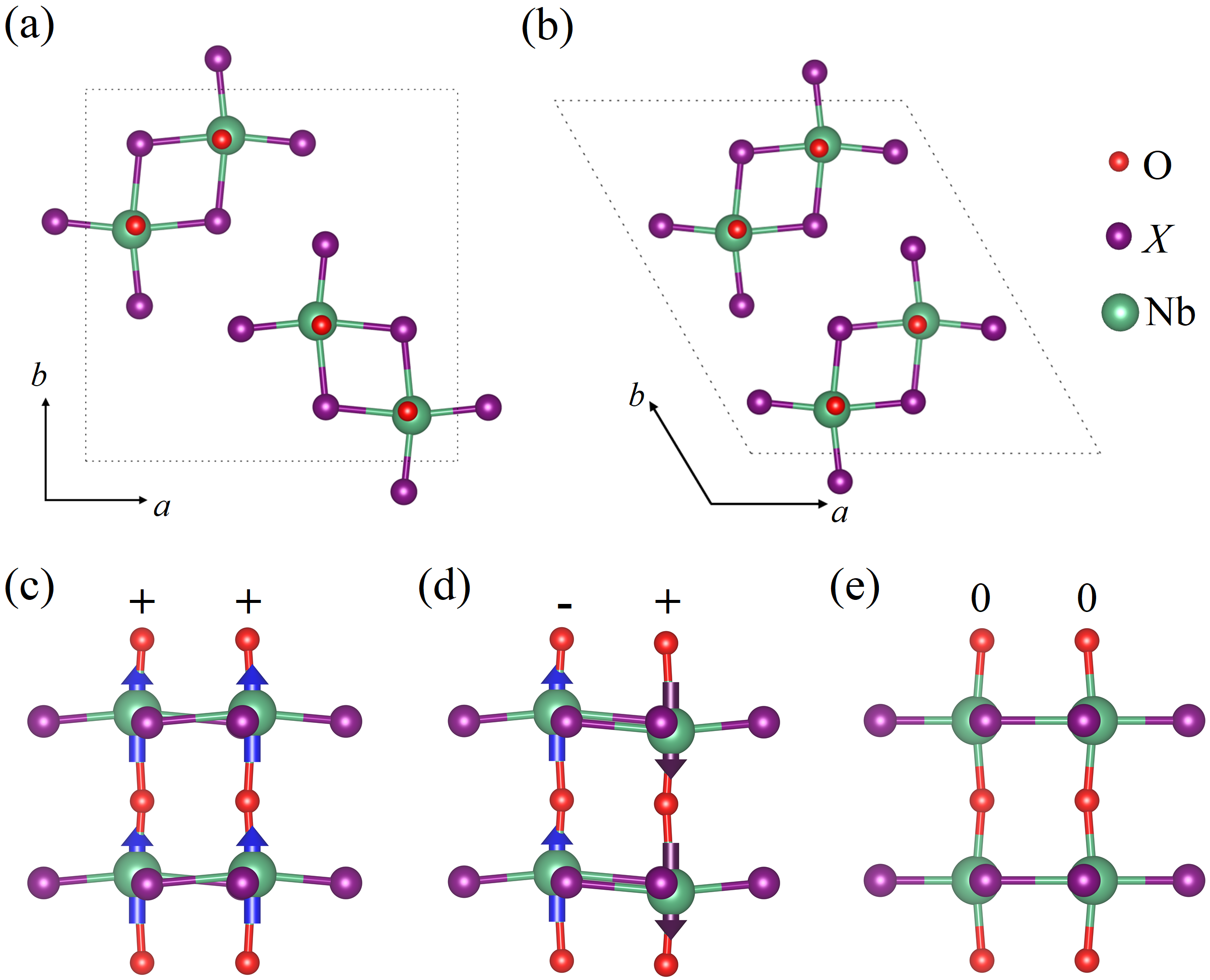}
\caption{Crystal structures of the NbO$X_3$ family. (a) and (b) are the top views of tetragonal and monoclinic lattices, respectively. (c)-(e) are 1D double-chains of FE, AFE, and paraelectric (PE) states, respectively. The symbols $+$/$-$ indicate the directions of polarizations. The blue/purple arrows represent $+$/$-$ polar direction along the $c$ axis.}
\label{fig1}
\end{figure}

Such as WO$X_4$ and VOI$_2$ \cite{Lin2019,Ding2020}, NbO$X_3$ series belong to the family of transition metal oxyhalides, which can be synthesized in experiments from reactions of niobium oxide and halogen at high temperatures \cite{Strobele2002,Hartwig2008a,Sands1959}. The experimental structures of NbO$X_3$ family are shown in Figs.~\ref{fig1}(a) and \ref{fig1}(b). In particular, NbOCl$_3$ and NbOBr$_3$ own the tetragonal phase, whereas NbOI$_3$ owns the monoclinic one, both of which consist of a pair of 1D double-chains in each u.c. The double-chains are coupled via vdW interaction, forming quasi-1D structures. Within each chain, each niobium is within an octahedral  cage, which is formed by four planar halogen ions and two apical oxygen ions. Then two parallel NbO$_2X_4$ chains are linked into a dimer via a shared edge of two halogen ions, forming a unique double-chain. In contrast, the vdW structure in WO$X_4$ is consisted of single chains.

Since the nominal value of Nb is $+5$ here, its $4d^0$ electron configuration will be a primary driving force for polar instability. According to the experimental structure, all Nb ions exhibit relative displacements from the center of octahedra along the Nb-O-Nb 1D chain direction, leading to local dipoles. In the previous study, two polar structures of 1D NbO$X_3$ double-chain have been explored \cite{Zhang2021a}. The FE and AFE structures are shown in Figs.~\ref{fig1}(c) and \ref{fig1}(d), where the dipoles in two parallel chains are parallel and antiparallel, respectively. For reference, the centrosymmetric PE phase is shown in Fig.~\ref{fig1}(e), without the displacements of Nb ions along the chain direction. Their calculations found the 1D FE phase is energetically more stable than the AFE one ($0.1$/$1.0$/$1.1$ meV/Nb for Cl/Br/I)  \cite{Zhang2021a}.

\begin{table*}
\centering
\setlength{\tabcolsep}{0.022\textwidth}
\caption{DFT results of the NbO$X_3$ family, in comparison with the experimental (EXP) results. The lattice constants ($a$, $b$, and $c$) are in units of angstroms. The volume of unit cell ($V$) is in units of \AA$^3$. The energy differences $\Delta E$ are in units of meV/u.c., and the energy of monoclinic FE phase is taken as a reference. The lowest energy ones are emphasized in bold. The band gap is in units of eV.}
\begin{tabular*}{\textwidth}{lccccccccc}
\hline \hline
Material & Crystal & Phase & Space group & $a$ & $b$ & $c$ & $\Delta E$ & $V$ & Gap\\
\hline
\multirow{9}*{NbOCl$_3$} & \multirow{4}*{Monoclinic} 
	& PE & $I2/m$ & 11.020 & 12.705 & 3.817 & 136.68 & 466.88 & 1.35\\	
& 	& FE & $I2$ & 10.953 & 12.645 & 3.920 & 0 & 473.47 & 2.75\\	
&	& AFE-1 & $P2/b$ & 10.928 & 12.637 & 3.921 & 1.32 & 472.78 & 2.77\\
&   & AFE-2 & $P2_1/b$ & 10.927 & 12.632 & 3.921 & 1.91 & 472.78 & 2.77\\
&	& AFE-3 & $C\bar1$ & 10.933 & 12.630 & 3.922 & 1.06 & 473.03 & 2.75\\
\cline{2-10}
& \multirow{5}*{Tetragonal} & PE & $P4_2/mnm$ & 11.048 & & 3.817 & 126.50 & 465.81 & 2.44\\
&	& {\bf FE} & $P4_2/nm$ & 10.982 & & 3.918 & $-$6.10 & 473.26 & 2.83\\	
&	& AFE-1 & $P\bar{4}2_1m$ & 10.988 & & 3.919 & $-$4.86 & 472.67 & 2.83\\
&	& AFE-2 & $P2_1/C$ & 10.982 & & 3.919 & $-$4.11 & 472.52 & 2.84\\
&	& EXP & $P\bar{4}2_{1}m$ & 10.896 & & 3.948\\
\hline						
\multirow{9}*{NbOBr$_3$} & \multirow{5}*{Monoclinic} & PE & $I2/m$ & 11.744 & 13.553 & 3.832 & 92.49 & 529.81 & 1.78\\	
& 	& FE & $I2$ & 11.703 & 13.531 & 3.914 & 0 & 536.99 & 1.95\\	
&	& AFE-1 & $P2/b$ & 11.673 & 13.511 & 3.911 & 1.34 & 535.85 & 1.99\\
&	& AFE-2 & $P2_1/b$ & 11.660 & 13.510 & 3.909 & 4.29 & 535.84 &  1.93\\	
&	& AFE-3 & $C\bar1$ & 11.662 & 13.502 & 3.910 & 5.3 & 534.64 & 1.93\\
\cline{2-10}
& \multirow{5}*{Tetragonal} & PE & $P4_{2}/mnm$ & 11.756 & & 3.817 & 88.15 & 529.30 & 1.71\\
&	& {\bf FE} & $P4_2/nm$ & 11.699 & & 3.910 & $-$1.63 & 535.18 & 1.93\\
&	& AFE-1 & $P\bar{4}2_1m$ & 11.606 & & 3.905 & $-$0.53 & 535.25 & 1.91\\
&	& AFE-2 & $P2_1/{C}$ & 11.707 & & 3.908 & 3.24 & 535.19 & 1.91\\
&	& EXP & $P\bar{4}2_1m$ & 11.636 & & 3.953\\
\hline						
\multirow{10}*{NbOI$_3$} & \multirow{6}*{Monoclinic} & PE & $I2/m$ & 12.778 & 14.795 & 3.878 & 46.32 & 633.53 & 0.61\\
& 	& {\bf FE} & $I2$ & 12.738 & 14.764 & 3.939 & 0 & 640.00 & 0.76\\	
&	& AFE-1 & $P2/b$ & 12.744 & 14.759 & 3.938 & 3.14 & 638.73 & 0.76\\
&	& AFE-2 & $P2_1/b$ & 12.730 & 14.762 & 3.937 & 6.12 & 639.60 & 0.74\\
&	& AFE-3 & $C\bar1$ & 12.726 & 14.756 & 3.936 & 6.55 & 638.85 & 0.78\\
& 	& EXP & $I2$ & 12.602 & 14.624 & 3.991 & & &\\
\cline{2-10}
& \multirow{4}*{Tetragonal} & PE & $P4_{2}/mnm$ & 12.768 & & 3.876 & 59.53 & 631.88 & 0.52\\
&	& FE & $P4_2/nm$ & 12.723 & & 3.937 & 16.45 & 637.22 & 0.70\\
&	& AFE-1 & $P\bar{4}2_1m$ & 12.709 & & 3.936 & 19.05 & 635.70 & 0.67\\
&	& AFE-2 & $P2_1/C$ & 12.729 & & 3.934 & 21.76 & 637.24 & 0.65\\
\hline \hline
\end{tabular*}
\label{Table}
\end{table*}

In the following, we will explore the polarity of NbO$X_3$ beyond the simple 1D double-chain, but focus on the more realistic bulks. Since its unit cell contains a pair of 1D double-chains, there are more possible configurations, including: (1) the bulk FE states with all Nb dipoles pointing to the same directions ($++$, $++$); (2) the bulk AFE phases (named as AFE-1) consisted of 1D FE double-chains with opposite dipoles ($++$, $--$); (3) the AFE-2 ($+-$, $-+$) and AFE-3 ($+-$, $+-$) structures consisted of 1D AFE double-chains with different stacking modes, as shown in Fig.~S1 in the Supplemental Materials (SM) \cite{Supp}. For the tetragonal crystals, the AFE-2 and AFE-3 structures are identical due to symmetry.

Our calculated results are summarized in Table~\ref{Table}. It is clear that all DFT optimized lattice constants are close to the corresponding experimental values. For NbOI$_3$, the ground state is indeed in the monoclinic FE phase, whereas the tetragonal phases are definitely higher in energy. In contrast, the tetragonal structures of NbOCl$_3$ and NbOBr$_3$ are more stable than their monoclinic counterparts. Both these results agree with experimental structures \cite{Strobele2002,Hartwig2008a}.

Surprisingly, it is noted that the calculated ground states of both NbOCl$_3$ and NbOBr$_3$ are tetragonally FE (teFE) phases, which are slightly lower in energy ($\sim1$ meV/u.c.) than the tetragonal AFE-1 (teAFE-1) expected from the experimental structure \cite{Strobele2002,Hartwig2008a}. Such tiny energy differences between the teFE and the teAFE-1 phases are reasonable considering the weak interchain vdW interaction. Since the DFT calculation is for the zero-temperature ground state. Then the experimental teAFE-1 phase may exist at an ambient condition due to the thermal excitation or entropy effect. In fact, such proximate energies and structures may allow the phase coexistence (and MPB) in real materials.

To verify this hypothesis, we simulated the powder XRD patterns of tetragonal NbOCl$_3$ using vesta \cite{Momma2011}, based on our optimized structures. As shown in Fig.~S2 in the SM \cite{Supp}, there are indeed some characteristic peaks e.g. (201) and (401), to distinguish the PE and AFE-1 phases, as used in the early XRD experiment to identify the AFE-1 phase \cite{Strobele2002}. However, the XRD patterns are extremely similar between the FE and the AFE-1 phases, without any distinct peak in the measuring region. It was possible the reason why the early XRD measurement did not identify the teFE phases.

\subsection{Dynamic stability and evolution under pressure}
Taking the tetragonal NbOCl$_3$ and monoclinic NbOI$_3$ as examples, the phonon spectra and symmetry analyses are performed to investigate the dynamic stability and the evolution of phases. 

\begin{figure}
\centering
\includegraphics[width=0.48\textwidth]{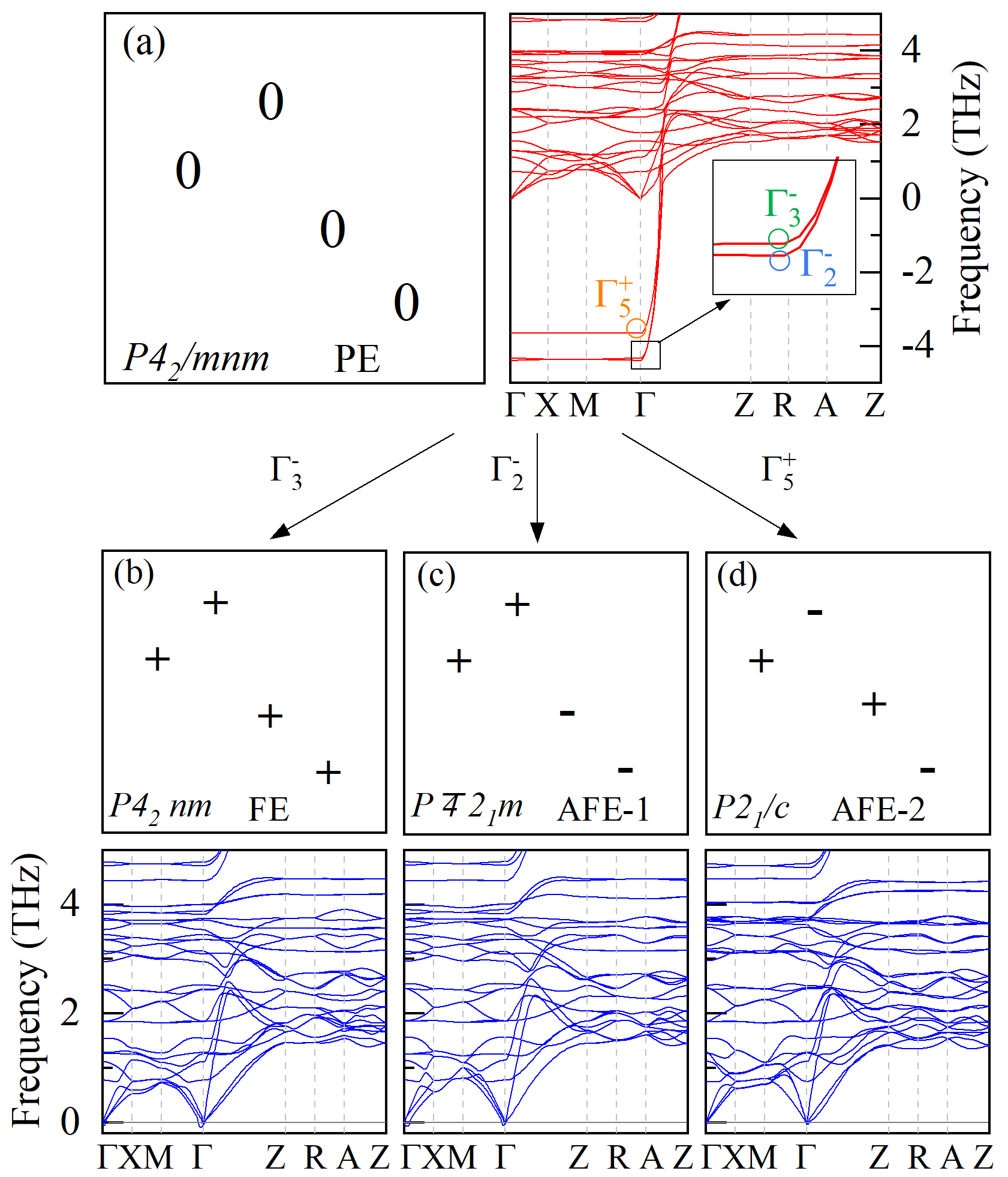}
\caption{Distortion modes and phonon spectra of tetragonal NbOCl$_3$. (a) PE; (b) FE; (c) AFE-1; (d) AFE-2. The inset of (a): the magnified view to distinguish the almost degenerate $\Gamma_3^-$ and $\Gamma_2^-$.}
\label{fig2}
\end{figure}

As shown in Fig.~\ref{fig2}(a), there are four branches of imaginary frequencies crossing the center and edge of the BZ for the phonon spectrum of tetragonal PE NbOCl$_3$. Specifically, the $\Gamma_3^-$ mode leads to the FE phase [Fig.~\ref{fig2}(b)], and the $\Gamma_2^-$ mode leads to the AFE-1 phase [Fig.~\ref{fig2}(c)]. As aforementioned, the difference between the FE and the AFE-1 is only the inter-chain coupling: parallel vs anti-parallel, whereas their intra-chain distortions are identical. Thus, these $\Gamma_3^-$ and $\Gamma_2^-$ modes are almost degenerate in imaginary frequencies. Another double-degenerate $\Gamma_5^+$ mode leads to AFE-2 (or equivalent AFE-3) bulk structure [Fig.~\ref{fig2}(d)]. There is no any imaginary frequency in the phonon spectra of FE, AFE-1, and AFE-2 phases, indicating their dynamic stabilities.

\begin{figure}
\includegraphics[width=0.48\textwidth]{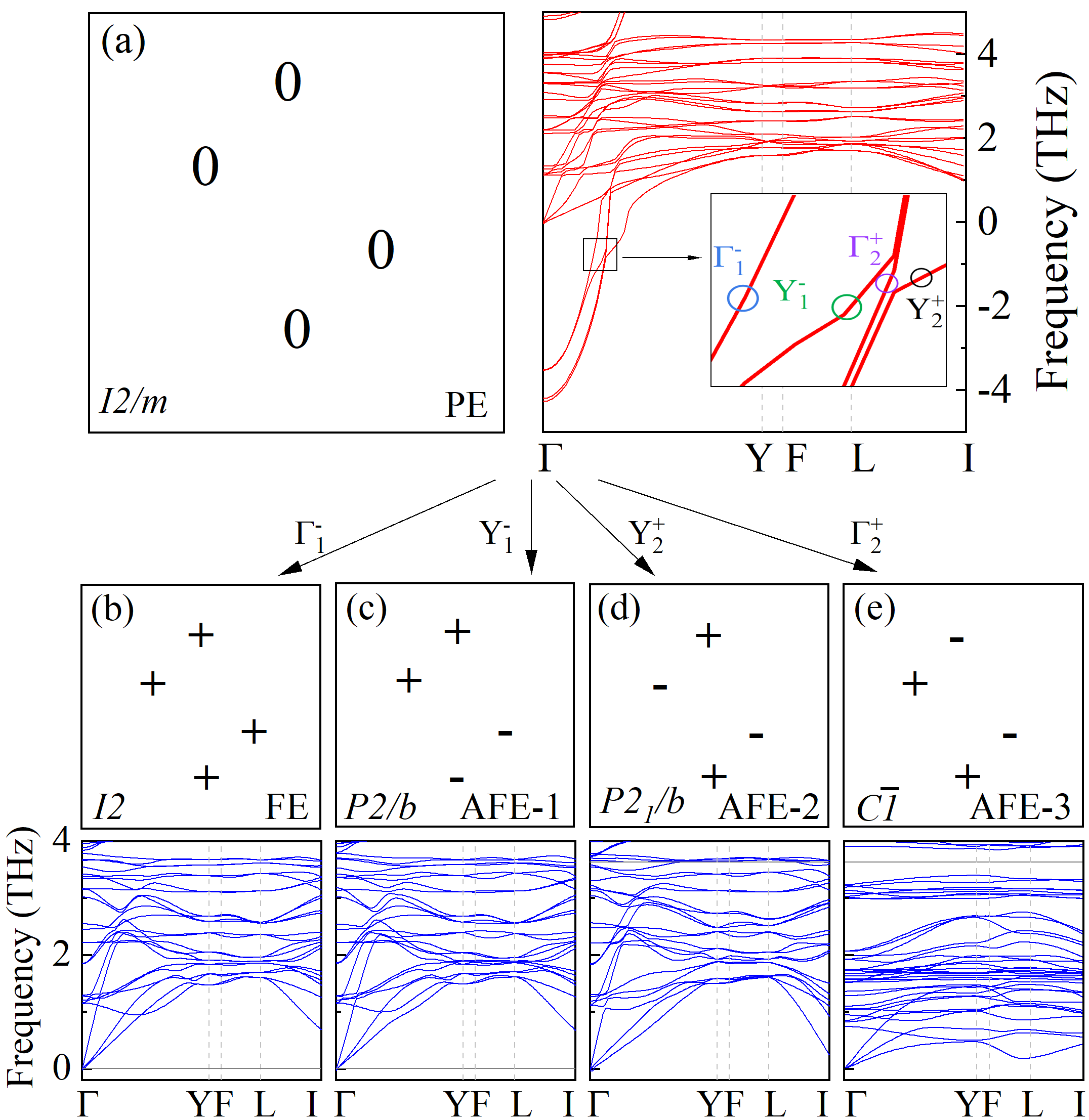}
\caption{Distortion modes and phonon spectra of monoclinic NbOI$_3$. (a) PE; (b) FE; (c) AFE-1; (d) AFE-2; (e) AFE-3. The inset of (a): the magnified view to distinguish the almost degenerate $\Gamma_1^-$ and $Y_1^-$, as well as $\Gamma_2^+$ and $Y_2^+$.}
\label{fig3}
\end{figure}

Similar analyses are also performed for monoclinic NbOI$_3$. As shown in Fig.~\ref{fig3}(a), there are also four imaginary frequencies located at the center of the BZ in the phonon spectrum of the PE phase. The higher two modes ($\Gamma_1^-$ and $Y_1^-$) are almost degenerate, which lead to the FE and AFE-1 phases, respectively. The lower two modes ($Y_2^+$ and $\Gamma_2^+$) are almost degenerate too, which lead to the AFE-2 and AFE-3 phases, respectively. Also, for all these four FE/AFE phases, they are dynamicly stable since no imaginary frequency exist in their phonon spectra, as shown in Figs.~\ref{fig3}(b) and \ref{fig3}(e).

More phonon spectra of monoclinic NbOCl$_3$, tetragonal NbOI$_3$, and monoclinic/tetragonal NbOBr$_3$ can be found in the SM (Figs.~S3-S6) \cite{Supp}. Their physical scenarios are quite similar.

According to above analyses, there are seven possible stable structures for each NbO$X_3$ crystal. It can be seen from Table~\ref{Table} that their energies are rather proximate, especially for NbOCl$_3$ and NbOBr$_3$. These small energy differences (on the order of $1$ meV/Nb) make it highly possible to tune these phases using external stimulations. For example, a hydrostatic pressure may drive the monoclinic ones to tetragonal phases, since the tetragonal ones always own smaller volumes, as compared in Table~\ref{Table}. Since the ground states of NbOCl$_3$ and NbOBr$_3$ are already the tetragonal phases, this pressure driven phase transition can only occur in NbOI$_3$ whose ground state is monoclinic.

\begin{figure}
\includegraphics[width=0.48\textwidth]{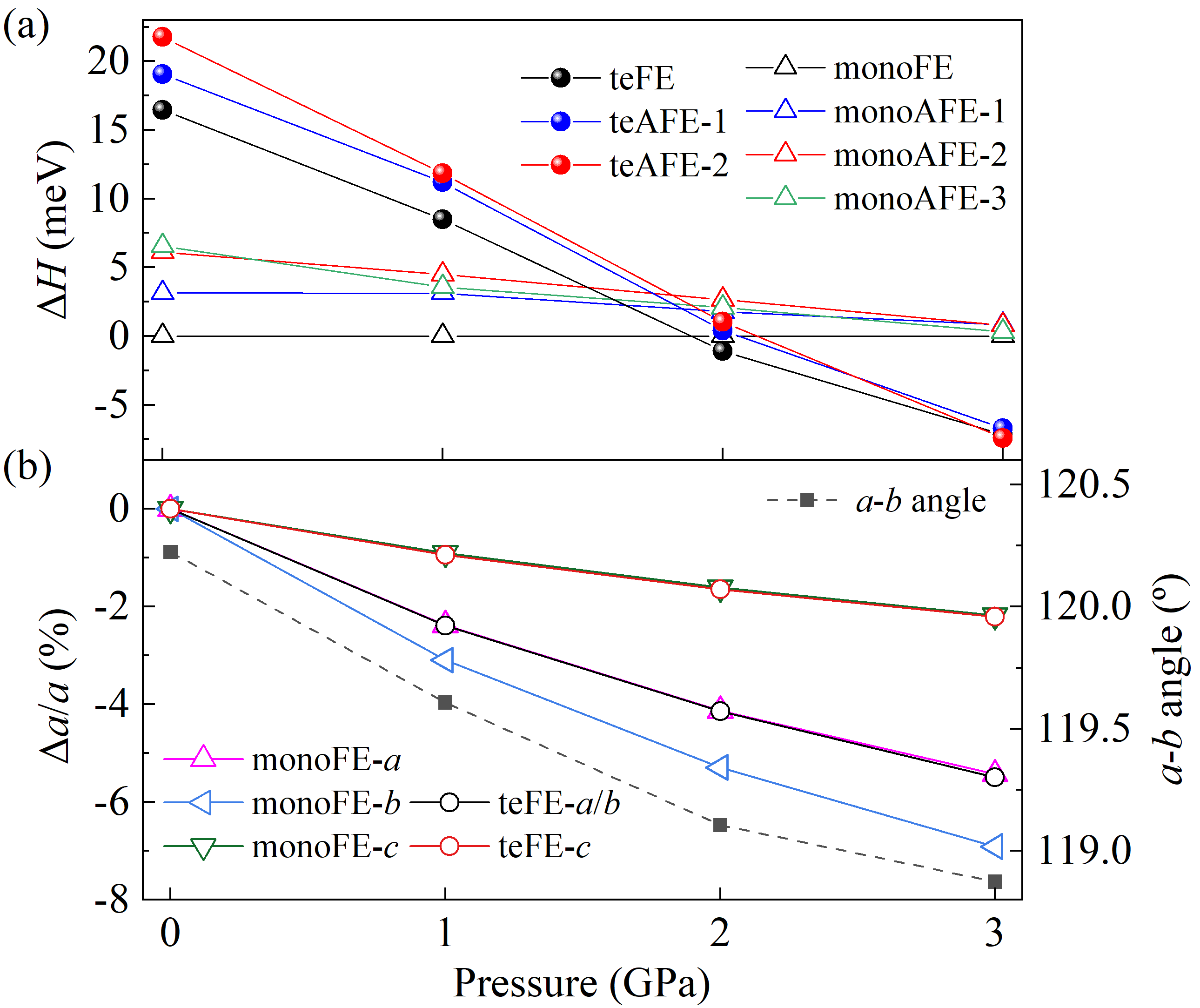}
\caption{(a) The relative enthalpies (per u.c.) of various phases of NbOI$_3$ as a function hydrostatic pressure. The monoclinic FE phase is taken as the reference. (b) The relative changes of lattice constants (left) and lattice angle between the $a$-$b$ axes (right) under pressure. Only the FE phase is illustrated as an example, whereas the tendencies of other AFE phases are similar.}
\label{fig4}
\end{figure}

To investigate possible structural phase transition, the structures of NbOI$_3$ are optimized under hydrostatic pressures. The enthalpies are compared to determined the possible ground states under the pressure, as shown in Fig.~\ref{fig4}(a). It is clear that NbOI$_3$ will turn to be tetragonal when pressure is larger than $2$ GPa. The evolution of lattice constants are shown in Fig.~\ref{fig4}(b).  Under expectation, the in-plane lattice constants shrink much larger than the out-of-plane one, due to the anisotropic stiffness.  The vdW interaction between chains make the material softer on the $ab$ plane. For the monoclinic phase, the $b$ axis is even more softer than the $a$ axis, in agreement with the intuitive view of its structure [Fig.~\ref{fig1}(b)]. Meanwhile, the lattice angle between the $a$-$b$ axes of the monoclinic phase also decreases with pressure, also shown in Fig.~\ref{fig4}(b).

\subsection{Ferroelectricity and piezoelectricity}
\begin{figure}
\includegraphics[width=0.48\textwidth]{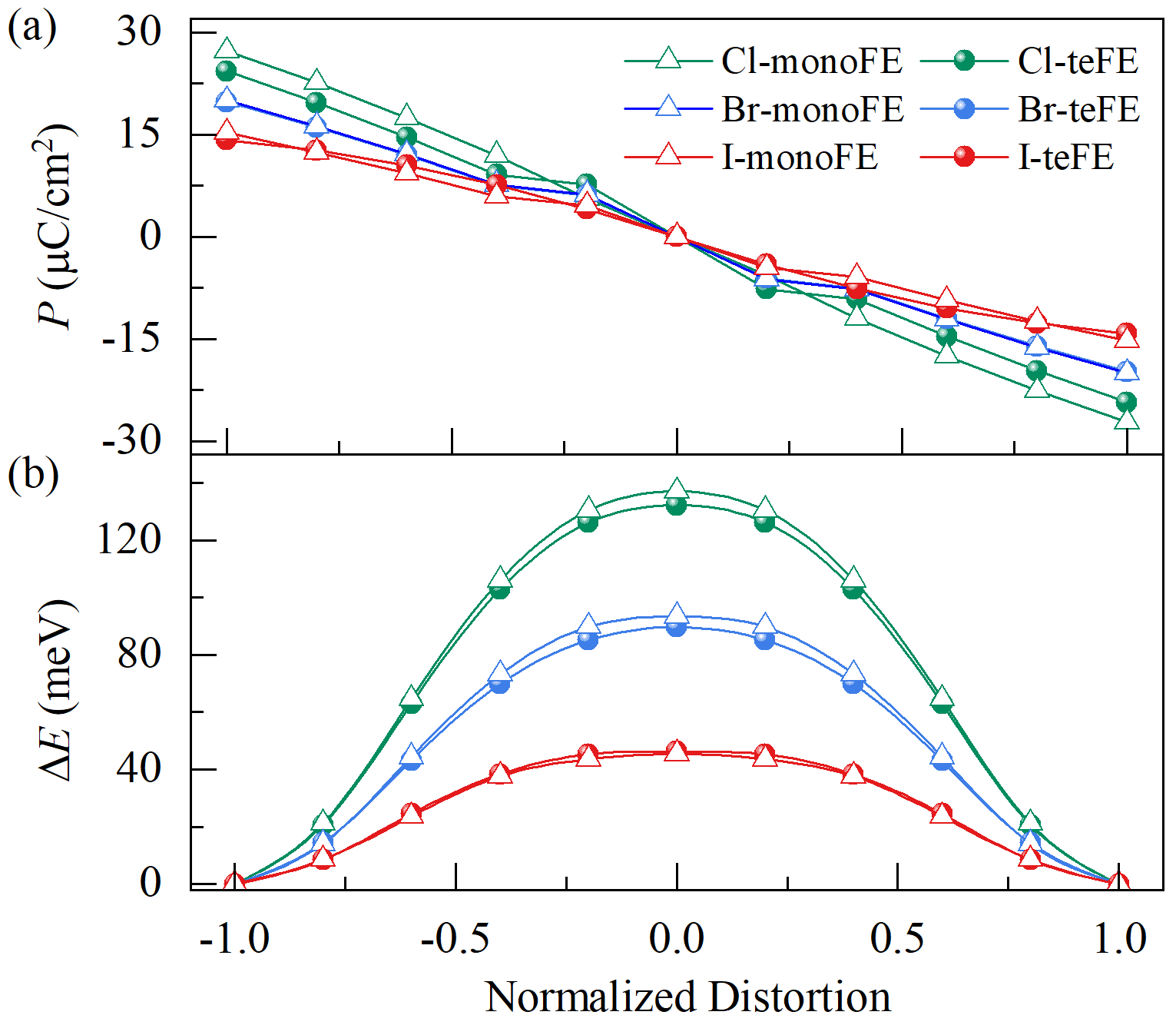}
\caption{Ferroelectric switching of  NbO$X_3$ bulks. (a) Polarization and (b) energy barriers (per u.c.) as a function of the FE distortion intensity. Two end states ($\pm1$) denote the optimized FE states ($\pm P$), and the middle one is the optimized PE states. Other intermiddle points are obtained using structural interpolation.}
\label{fig5}
\end{figure}

The polarization is a key physical quantity of a FE material. Here the spontaneous polarizations of NbOCl$_3$/NbOBr$_3$/NbOI$_3$ are estimated to be $24.3$/$19.8$/$14.2$ $\mu$C/cm$^2$ for tetragonal FE phases, and $27.2$/$20.0$/$15.27$ $\mu$C/cm$^2$ for monoclinic FE phases, respectively, as shown in Fig.~\ref{fig5}(a). Due to the quasi-1D characteristic, these values are almost identical for those meta-stable structures, i.e., monoclinic NbOCl$_3$/NbOBr$_3$ and tetragonal NbOI$_3$, as well as the single double-chains \cite{Zhang2021a}. The deceasing polarization from $X$=Cl to I is mainly due to the increasing volume. Luckily, all these polarizations are moderate, comparable to that of the commercially used perovskite BaTiO$_3$ ($\sim20-25$ $\mu$C/cm$^2$). 

Besides the polarization, the coercive field is also an essential property of a FE system, which is, in principle, related to the energy cost of FE switching barrier. Seven possible flip paths are considered, in which the transition states include PE, FE, AFE-1, AFE-2 and AFE-3 phases (for monoclinic structures only), as shown in Fig. S7 of the SM \cite{Supp}. The calculation results show that the FE-PE flip path has the lowest potential barrier for all three cases. As shown in Fig.~\ref{fig5}(b), the estimated barrier of NbOCl$_3$/NbOBr$_3$/NbOI$_3$ is $132$/$89$/$46$ meV/u.c. (i.e., $33$/$22$/$12$ meV/f.u.) for tetragonal FE phases, and $137$/$93$/$45$ meV/u.c. (i.e., $34$/$23$/$11$ meV/f.u.) for monoclinic NbOI$_3$, which is comparable to that of BaTiO$_3$ ($\sim34$ meV/f.u.) \cite{Lin2019}. Noting here these barriers can only be considered as theoretical upper limits, whereas other switching pathes may exist which can further reduce the barriers.

The ferroelectricity of NbO$X_3$ can be described by a classical polarization model based on the Landau-Ginzburg-Devonshire formula \cite{PRL.LLF}:
\begin{equation}
E=\frac{\alpha}{2}P^{2}_{1}+\frac{\beta}{4}P^{4}_{1}+\frac{\gamma}{6}P^{6}_{1}+\frac{\alpha}{2}P^{2}_{2}+\frac{\beta}{4}P^{4}_{2}+\frac{\gamma}{6}P^{6}_{2}+\frac{\delta}{2}(P^{2}_{1}-P^{2}_{2})^{2},
\label{key}
\end{equation}
where $P_1$ and $P_2$ denote the (local) polarization from double-chains 1 and 2, respectively, and the last item corresponds to the interchain coupling. All these coefficients can be obtained by fitting the polarization-potential curves. The coefficients of the NbO$X_3$ FE phases and the fitting polarization-energy barrier curves are shown as Table~S1 and Fig.~S8 of the SM \cite{Supp}. The calculation results show that the values of $\delta$ are substantially smaller than that of $\alpha$, indicating the inter-chain coupling is weak.

Besides ferroelectricity, piezoelectricity is another important physical property for FE materials. Crystallographic symmetry analysis shows that the elastic matrix $C$'s have 7 and 13 independent non-zero coefficients for the tetragonal and monoclinic lattices respectively \cite{Mouhat2014}. The piezoelectric stress tensor matrix $e$ of tetragonal (monoclinic) crystal has four- (five-) independent matrix elements \cite{DeJong2015}. Then the piezoelectric strain coefficient calculation formula can be expressed as:
\begin{equation}
d_{ij}=\Sigma_{k=1}^{6}e_{ik}S_{kj},
\label{key}
\end{equation}
where $S$ denotes the tensor of the elastic compliance coefficient ($S=C^{-1}$). The key piezoelectric coefficient $d_{33}$ represents the piezoelectric constant  when the polarization direction is the same as the stress direction. Here the calculated piezoelectric $d_{33}$'s are $34.6$/$36.9$/$60.1$ pC/N for FE tetragonal NbOCl$_3$/NbOBr$_3$ and monoclinic NbOI$_3$ phases, respectively. All these $d_{33}$'s are larger than those of quasi-1D WO$X_4$ (where their $d_{33}$=$27.3$/$27.7$/$39.8$ pC/N for $X$ = F/Cl/Br) \cite{Lin2019}, as well as the conventional piezoelectric ZnO ($d_{33}$= $12.3$ pC/N) \cite{Wu2005}. Note that the theoretical $d_{33}$'s should be considered to be lower limits, whereas in real materials the experimental $d_{33}$'s can be enhanced significantly by many other factors such as MPB and domain textures. More details of the elastic tensor $C$, the piezoelectric tensor $e$, and the piezoelectric strain tensors $d$ are summarized in Table~S2 of the SM \cite{Supp}.

The thermal stablility has also been checked. With a $1\times1\times5$ supercell, our AIMD simulations at $500$ K for $5$ ps show robust structures for both the tetragonal NbO$X_3$ ($X$=Cl/Br) and monoclinic NbOI$_3$, as shown in Fig.~S9 of SM \cite{Supp}. 

Furthermore, the SOC has also been tested on the NbOI$_3$ ground state, which owns the heaviest atom iodine. However, the lattice constants, band gap, and polarization are almost unchanged with SOC, as shown in Table~S3 of the SM \cite{Supp}. Physically, the impact of SOC on NbOCl$_3$ and NbOBr$_3$ should be even weaker. Such negligible SOC effects are reasonable, since all of them are band insulators with fully occupied/empty bands.

\section{Conclusion}
To summarize, the quasi-1D NbO$X_3$ bulks have been systematically studied, going beyond the individual double-chain scenario but considering more relastic fact of inter-chain interactions. From the parent PE phases, seven possible FE or AFE states are derived according to the symmetry analyses, whose dynamic stabilities are further confirmed by phonon spectra. The FE and AFE phases are energetically proximate, which may allow the coexisting in real materials at ambient conditions. And for NbOI$_3$ the monoclinic to tetragonal structural transition can be triggered by a small hydrostatic pressure. The polarities of FE NbO$X_3$ are prominent, with considerable spontaneous polarizations, moderate switching barriers, and large piezoelectric $d_{33}$ coefficients, which allow promising potentials for applications.

\begin{acknowledgements}
This work was supported by National Natural Science Foundation of China (Grants No. 11834002 and No. 12104089), Natural Science Foundation of Jiangsu Province (Grant No. BK20200345) the Fundamental Research Funds 357 for the Central Universities. We acknowledge the computational source  from the Big Data Center of Southeast University.
\end{acknowledgements}

\bibliography{NbOX_ref}
\bibliographystyle{apsrev4-2}
\end{document}